\documentclass[aps,reprint,amsmath,amssymb]{revtex4-1}

\usepackage{graphicx}
\usepackage{dcolumn}
\usepackage{bm}

\usepackage{amsmath} 
\usepackage{amssymb}
\usepackage{amsfonts}
\usepackage{upgreek} 

\usepackage{xspace}
\usepackage{ifthen} 
\newboolean{articletitles}
\setboolean{articletitles}{true} 

\newboolean{uprightparticles}
\setboolean{uprightparticles}{false} 

\newboolean{inbibliography}
\setboolean{inbibliography}{false} 

\usepackage{mciteplus}

\usepackage{xspace} 
\usepackage{upgreek}

\def\lhcb {\mbox{LHCb}\xspace}
\def\atlas  {\mbox{ATLAS}\xspace}
\def\cms    {\mbox{CMS}\xspace}

\def\MagUp {\mbox{\em Mag\kern -0.05em Up}\xspace}

\ifthenelse{\boolean{uprightparticles}}%
{

 \def\Pmu         {\ensuremath{\upmu}\xspace}

 \def\Ppi         {\ensuremath{\uppi}\xspace}

 \def\Ppsi        {\ensuremath{\uppsi}\xspace}

 \def\PDelta      {\ensuremath{\Delta}\xspace}                 
 \def\PXi      {\ensuremath{\Xi}\xspace}                 
 \def\PLambda      {\ensuremath{\Lambda}\xspace}                 
 \def\PSigma      {\ensuremath{\Sigma}\xspace}                 
 \def\POmega      {\ensuremath{\Omega}\xspace}                 
 \def\PUpsilon      {\ensuremath{\Upsilon}\xspace}                 
 
 \def\PB      {\ensuremath{\mathrm{B}}\xspace}                 
                  
 \def\PD      {\ensuremath{\mathrm{D}}\xspace}

 \def\PJ      {\ensuremath{\mathrm{J}}\xspace}                 
 \def\PK      {\ensuremath{\mathrm{K}}\xspace}

 \def\Pb      {\ensuremath{\mathrm{b}}\xspace}                 
 \def\Pc      {\ensuremath{\mathrm{c}}\xspace}

 \def\Pi      {\ensuremath{\mathrm{i}}\xspace}

 \def\Pp      {\ensuremath{\mathrm{p}}\xspace}

 \def\Ps      {\ensuremath{\mathrm{s}}\xspace}

}
{

 \def\Pmu         {\ensuremath{\mu}\xspace}

 \def\Ppi         {\ensuremath{\pi}\xspace}

 \def\Ppsi        {\ensuremath{\psi}\xspace}                 
                  
 \mathchardef\PDelta="7101
 \mathchardef\PXi="7104
 \mathchardef\PLambda="7103
 \mathchardef\PSigma="7106
 \mathchardef\POmega="710A
 \mathchardef\PUpsilon="7107
                  
 \def\PB      {\ensuremath{B}\xspace}                 
                  
 \def\PD      {\ensuremath{D}\xspace}

 \def\PJ      {\ensuremath{J}\xspace}                 
 \def\PK      {\ensuremath{K}\xspace}

 \def\Pb      {\ensuremath{b}\xspace}                 
 \def\Pc      {\ensuremath{c}\xspace}

 \def\Pi      {\ensuremath{i}\xspace}

 \def\Pp      {\ensuremath{p}\xspace}

 \def\Ps      {\ensuremath{s}\xspace}

}

\makeatletter
\gdef\@ptsize{0} 
\makeatother

\DeclareRobustCommand{\optbar}[1]{\shortstack{{\miniscule (\rule[.5ex]{1.25em}{.18mm})}
  \\ [-.7ex] $#1$}}

\def\mup        {{\ensuremath{\Pmu^+}}\xspace}

\def\squark    {{\ensuremath{\Ps}}\xspace}

\def\cquark    {{\ensuremath{\Pc}}\xspace}

\def\bquark    {{\ensuremath{\Pb}}\xspace}

\def\pion   {{\ensuremath{\Ppi}}\xspace}
\def\piz    {{\ensuremath{\pion^0}}\xspace}

\def\pip    {{\ensuremath{\pion^+}}\xspace}
\def\pim    {{\ensuremath{\pion^-}}\xspace}

\def\kaon    {{\ensuremath{\PK}}\xspace}
  \def\Kbar    {{\kern 0.2em\overline{\kern -0.2em \PK}{}}\xspace}

\def\KorKbar    {\kern 0.18em\optbar{\kern -0.18em K}{}\xspace}

\def\Kp      {{\ensuremath{\kaon^+}}\xspace}
\def\Km      {{\ensuremath{\kaon^-}}\xspace}

  \def\Dbar    {{\kern 0.2em\overline{\kern -0.2em \PD}{}}\xspace}

\def\DorDbar    {\kern 0.18em\optbar{\kern -0.18em D}{}\xspace}

\def\B       {{\ensuremath{\PB}}\xspace}
\def\Bbar    {{\ensuremath{\kern 0.18em\overline{\kern -0.18em \PB}{}}}\xspace}

\def\BorBbar    {\kern 0.18em\optbar{\kern -0.18em B}{}\xspace}

\def\Bu      {{\ensuremath{\B^+}}\xspace}

\def\Bpm     {{\ensuremath{\B^\pm}}\xspace}

\def\Bd      {{\ensuremath{\B^0}}\xspace}
\def\Bs      {{\ensuremath{\B^0_\squark}}\xspace}

\def\jpsi     {{\ensuremath{{\PJ\mskip -3mu/\mskip -2mu\Ppsi\mskip 2mu}}}\xspace}

  \def\Y#1S{\ensuremath{\PUpsilon{(#1S)}}\xspace}

\def\proton      {{\ensuremath{\Pp}}\xspace}

\def\Lz          {{\ensuremath{\PLambda}}\xspace}
\def\Lbar        {{\ensuremath{\kern 0.1em\overline{\kern -0.1em\PLambda}}}\xspace}
\def\LorLbar    {\kern 0.18em\optbar{\kern -0.18em \PLambda}{}\xspace}

\def\Lb      {{\ensuremath{\Lz^0_\bquark}}\xspace}

\def\Lc      {{\ensuremath{\Lz^+_\cquark}}\xspace}

\def\to                 {\ensuremath{\rightarrow}\xspace}

\def\CP                {{\ensuremath{C\!P}}\xspace}

\def\AT#1     {\ensuremath{A_{\mathrm{T}}^{#1}}\xspace}           

\def\C#1      {\ensuremath{\mathcal{C}_{#1}}\xspace}                       
\def\Cp#1     {\ensuremath{\mathcal{C}_{#1}^{'}}\xspace}                    
\def\Ceff#1   {\ensuremath{\mathcal{C}_{#1}^{\mathrm{(eff)}}}\xspace}        
\def\Cpeff#1  {\ensuremath{\mathcal{C}_{#1}^{'\mathrm{(eff)}}}\xspace}       
\def\Ope#1    {\ensuremath{\mathcal{O}_{#1}}\xspace}                       
\def\Opep#1   {\ensuremath{\mathcal{O}_{#1}^{'}}\xspace}                    

\newcommand{\tev}{\ifthenelse{\boolean{inbibliography}}{\ensuremath{~T\kern -0.05em eV}\xspace}{\ensuremath{\mathrm{\,Te\kern -0.1em V}}}\xspace}
\newcommand{\gev}{\ensuremath{\mathrm{\,Ge\kern -0.1em V}}\xspace}
\newcommand{\mev}{\ensuremath{\mathrm{\,Me\kern -0.1em V}}\xspace}
\newcommand{\kev}{\ensuremath{\mathrm{\,ke\kern -0.1em V}}\xspace}
\newcommand{\ev}{\ensuremath{\mathrm{\,e\kern -0.1em V}}\xspace}
\newcommand{\gevc}{\ensuremath{{\mathrm{\,Ge\kern -0.1em V\!/}c}}\xspace}
\newcommand{\mevc}{\ensuremath{{\mathrm{\,Me\kern -0.1em V\!/}c}}\xspace}
\newcommand{\gevcc}{\ensuremath{{\mathrm{\,Ge\kern -0.1em V\!/}c^2}}\xspace}
\newcommand{\gevgevcccc}{\ensuremath{{\mathrm{\,Ge\kern -0.1em V^2\!/}c^4}}\xspace}
\newcommand{\mevcc}{\ensuremath{{\mathrm{\,Me\kern -0.1em V\!/}c^2}}\xspace}

\def\mum  {\ensuremath{{\,\upmu\mathrm{m}}}\xspace}

\def\mub{\ensuremath{{\mathrm{ \,\upmu b}}}\xspace}

\def\invpb {\ensuremath{\mbox{\,pb}^{-1}}\xspace}

\def\invfb   {\ensuremath{\mbox{\,fb}^{-1}}\xspace}

\def\gsim{{~\raise.15em\hbox{$>$}\kern-.85em
          \lower.35em\hbox{$\sim$}~}\xspace}
\def\lsim{{~\raise.15em\hbox{$<$}\kern-.85em
          \lower.35em\hbox{$\sim$}~}\xspace}

\def\pt         {\mbox{$p_{\mathrm{ T}}$}\xspace}

\def\evtgen     {\mbox{\textsc{EvtGen}}\xspace}

\def\geant      {\mbox{\textsc{Geant4}}\xspace}

\def\photos     {\mbox{\textsc{Photos}}\xspace}

\def\pythia     {\mbox{\textsc{Pythia}}\xspace}

\def\tell1  {TELL1\xspace}
\def\ukl1   {UKL1\xspace}

\begin{document}

\title{Review of recent LHCb results and expectations for Run II}

\author{A. Hicheur}
 \email{hicheur@if.ufrj.br}
\affiliation{%
 Instituto de F\'isica, Universidade Federal do Rio de Janeiro\\
 }%
\collaboration{On behalf of the \lhcb collaboration}
\begin{abstract}
As first Run II data acquisition has begun, it is useful to expose the pending questions by reviewing some of the most recent results obtained with Run I data analyses. Early results of the current data taking and middle-term prospects are also shown to illustrate the efficiency of the acquisition and analysis chain.
\end{abstract}

\maketitle

\section{Introduction}

The core of the \lhcb research program lies in the use of heavy flavour decays to probe the intervention of New Physics through the involvement of heavy new particles in loop decays. For this purpose, rare $b$-hadron decays proceeding through box or loop diagrams and in general any process in which a virtual heavy particle could intervene, come as natural grounds to seek for any deviation from Standard Model-based predictions.~Although the spectroscopy of heavy-flavoured hadrons and the study of new bound states might be considered as side-products of the main research lines, recent studies show that important results can emerge in this sector.

Beside low-energy hadron physics, the low transverse momentum threshold, the excellent vertex reconstruction, and the peculiar forward geometry of the \lhcb detector is allowing for complementary studies to what is performed with \atlas or \cms detectors, on areas such as QCD, Electroweak and Higgs physics, as well as searches for displaced vertices of heavy decaying particles.

The analyses exhibited in this paper are based on the full Run I (years 2011 and 2012) data sample, representing 1.0 and 2.0 \invfb of integrated luminosity at $7$ and $8$ TeV center-of-mass energies in $pp$ collisions, respectively. The early results and measurements of the first run II data at 13 TeV are based on samples of 4 to 6 \invpb, from the 50 ns proton bunch spacing run which occurred in early summer 2015.

\section{\lhcb detector and software}
\begin{figure}[htb]
\begin{center}
\includegraphics[width=0.5\textwidth]{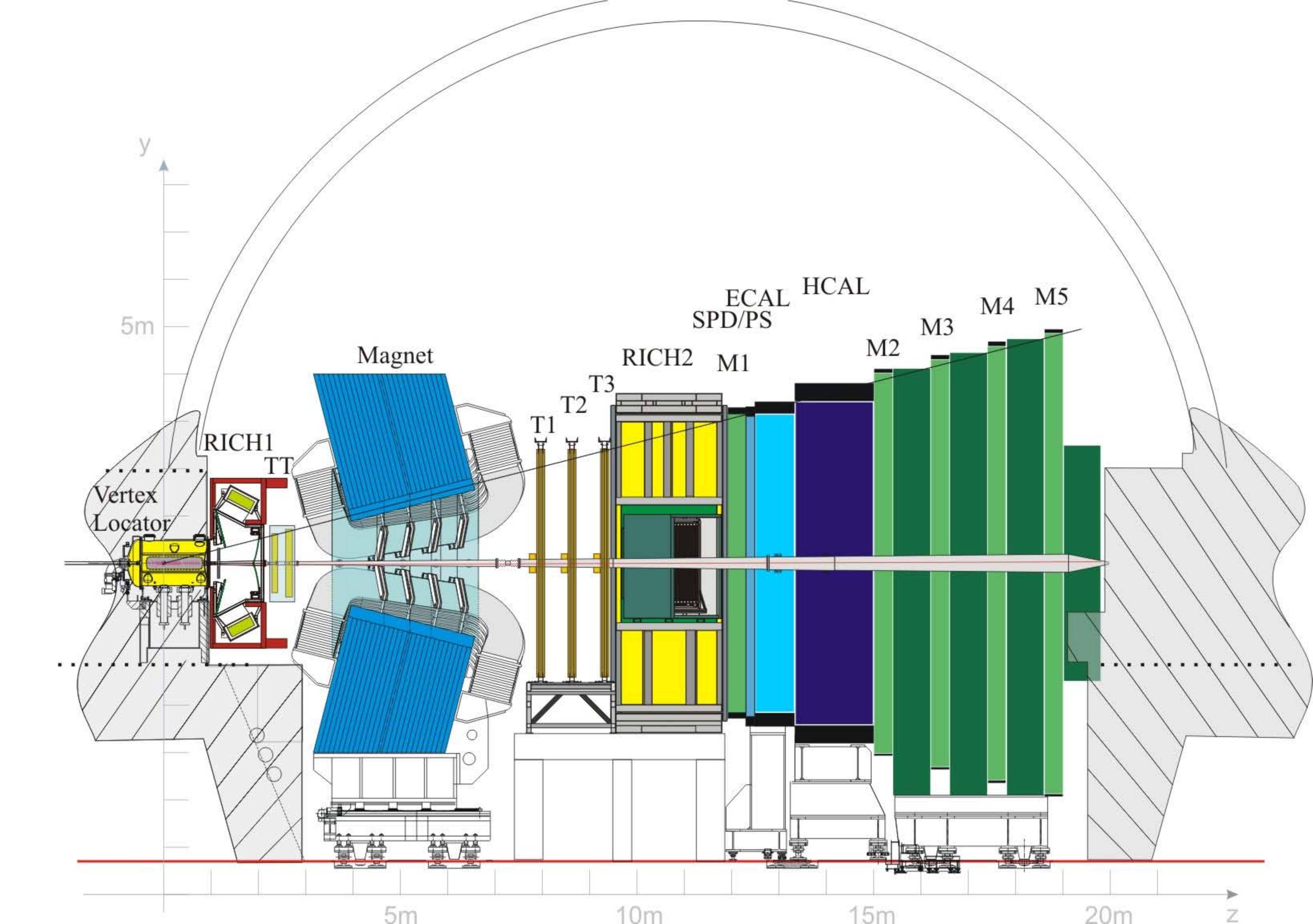}
\end{center}
\caption{Global view of the \lhcb detector.}
\label{Fig:lhcb_layout}
\end{figure}

The \lhcb detector~\cite{Alves:2008td}, Fig.\ref{Fig:lhcb_layout}, is a single-arm forward spectrometer covering the \mbox{pseudorapidity} range $2<\eta <5$, specially designed for the study of particles containing $b$ or $c$ quarks.~The detector includes a high precision tracking system, consisting of a silicon-strip vertex detector surrounding the $pp$ interaction region, a large-area silicon-strip detector located upstream of a dipole magnet with a bending power of about $4{\rm\,Tm}$, and three stations of silicon-strip detectors and straw drift tubes
 placed downstream.~The combined tracking system has momentum resolution $\Delta p/p$ that varies from 0.4\% at 5\gevc to 0.6\% at 100\gevc, and impact parameter~(IP) resolution of 20\mum for tracks with high transverse momentum.~Charged hadrons are identified using two ring-imaging Cherenkov detectors (RICH) \cite{RichPerf}. Photon, electron and hadron candidates are identified by a calorimeter system consisting of scintillating-pad and preshower detectors, an electromagnetic calorimeter and a hadronic calorimeter.~Muons are identified by a system composed of alternating layers of iron and multiwire proportional chambers. 

\begin{figure}[htb]
\begin{center}
\includegraphics[width=0.3\textwidth]{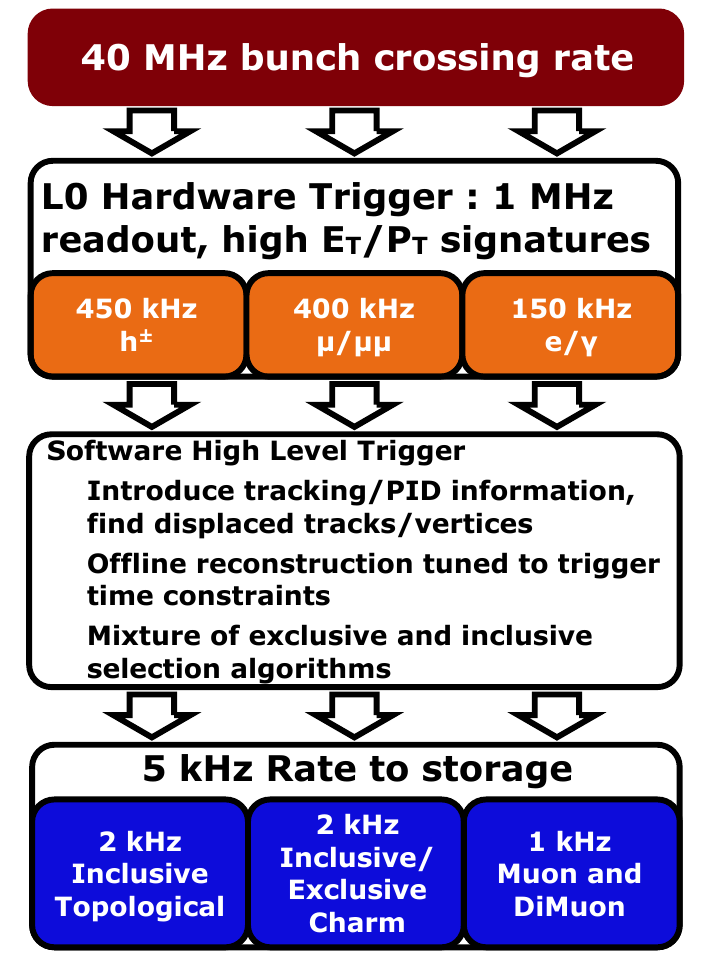}
\end{center}
\caption{\lhcb data acquisition chain for Run I data.}
\label{Fig:lhcb_trigger_2012}
\end{figure}

The data acquisition chain is depicted in Fig.\ref{Fig:lhcb_trigger_2012}. The trigger~\cite{Aaij:2012me} consists of a hardware stage, based on information from the calorimeter and muon systems, followed by a software stage which applies a full event reconstruction.~Events triggered both on objects independent of the signal, and associated with the signal, are used. In the latter case, the transverse energy of the hadronic cluster is required to be at least 3.5\gev. The software trigger includes more sophisticated requirements and operations, in particular a multivariate algorithm to identify secondary vertices~\cite{Gligorov:2012qt}.

The analyses use simulated events generated by \pythia~8.1~\cite{Sjostrand:2008} with a specific \lhcb configuration~\cite{LHCb-PROC-2010-056}.  Decays of hadronic particles are described by \evtgen~\cite{Lange:2001uf} in which final state radiation is generated using \photos~\cite{Golonka:2005pn}. The interaction of the generated particles with the detector and its response are implemented using the \geant toolkit~\cite{Allison:2006ve,*Agostinelli:2002hh} as described in Ref.~\cite{LHCb-PROC-2011-006}.

\section{Brief reminder on quark couplings}
Because the eigenstates of the weak interaction are different from the mass eigenstates, the Cabibbo-Kobayashi-Maskawa (CKM) matrix \cite{CKM1,CKM2} involves couplings within and between quark generations:
\begin{equation}
\label{CKM_matrix}
 V_{CKM}  =
\left[ 
       \begin{array}{c c c} 
              V_{ud} &  V_{us} & V_{ub} \\
              V_{cd} &  V_{cs} & V_{cb} \\
    	      V_{td} &  V_{ts} & V_{tb} 
       \end{array} 
\right]      
\end{equation}
However, the magnitude of its elements is not evenly distributed: the modules of the diagonal elements (within-generation couplings) are one while the strength decreases as one departs from the diagonal. The weakest couplings are thus between the first and third generation, namely $V_{td}$ and $V_{ub}$.
The unitarity condition, $ V_{CKM} V_{CKM}^\dagger=\mathbf{1}$, imposes six independent relations, of which two are of particular interest since one of them the ``\Bd triangle'', $V_{ ud } V_{ ub }^*+V_{ cd } V_{ cb }^*+V_{ td } V_{ tb }^*=0$, involves a sum of elements of similar magnitude, and the second one, $V_{ us} V_{ ub }^*+V_{ cs } V_{ cb }^*+V_{ ts } V_{ tb }^*=0$, although being a unbalanced squeezed triangle, involves the angle intervening in the \Bs meson oscillations, $\beta_s=-\mathrm{arg}\left(\frac{-V_{ts}V_{tb}^*}{V_{cs}V_{cb}^*}\right)$.

For the \Bd triangle, the angle $\beta=\mathrm{arg}\left(\frac{-V_{cd}V_{cb}^*}{V_{td}V_{tb}^*}\right)$ is now measured at a very high precision even with \lhcb measurement alone \cite{LHCb-PAPER-2015-004}. The angle $\gamma=\mathrm{arg}\left(\frac{-V_{ud}V_{ub}^*}{V_{cd}V_{cb}^*}\right)$, however is much less known and is the subject of an extensive program involving $B\to DK$ decays \cite{LHCb-CONF-2014-004}. The module of the element $V_{ ub }$ intervening in the latter, can be probed through the study of the semileptonic decays $b\to u~\ell^-\overline\nu$. Finally, the sides $V_{ ts } V_{ tb }^*$ and $V_{ td } V_{ tb }^*$ are involved in rare $B$ decays through flavour changing neutral currents $b\to s,d$.

\section{Rare $b\to (s,d) \ell^+\ell^-$ decays}
\begin{figure}[t]
\begin{center}
\includegraphics[width=0.35\textwidth]{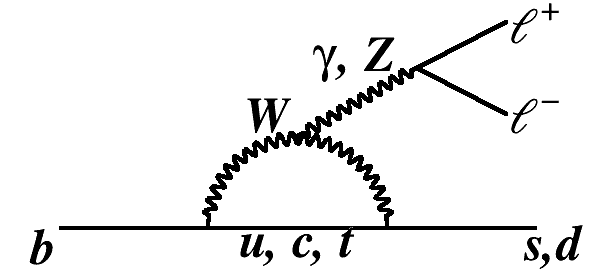}
\includegraphics[width=0.35\textwidth]{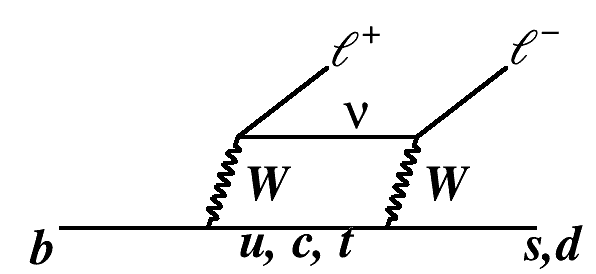}
\end{center}
\caption{Flavour changing electroweak $b\to s,d$ (top) loop and (bottom) box transitions.}
\label{Fig:btosll}
\end{figure}

$b\to (s,d) \ell^+\ell^-$ decays proceed in the Standard Model through flavour changing neutral currents (FCNC) involving loop or box diagrams as in Fig.\ref{Fig:btosll}. However, their dynamics could well be affected by the intervention of heavy virtual particles coming from a higher mass scale.
\lhcb has studied all of $\Bd\to K^{*0} \mu^+\mu^-$ \cite{LHCb-CONF-2015-002}, $\Bs\to \phi \mu^+\mu^-$ \cite{LHCb-PAPER-2015-023}, $\Lb\to \Lz \mu^+\mu^-$ \cite{LHCb-PAPER-2015-009}, for $b\to s \mu^+\mu^-$,  and $\Bu\to \pip \mu^+\mu^-$ \cite{LHCb-PAPER-2015-035}, for $b\to d \mu^+\mu^-$. A consistent mismatch between data and Standard Model(SM)-based predictions at low $q^2=m^2(\mu^+\mu^-)$ is exhibited for several observables in $b\to s \mu^+\mu^-$ decays. This the case for the decay $\Bd\to K^{*0} \mu^+\mu^-$ for which the variable $P_5^\prime=S_5/\sqrt{F_L(1-F_L)}$, which combines the coefficient $S_5$ of the angular distribution \cite{LHCb-CONF-2015-002} and the fraction of $K^{*0}$ longitudinal polarization $F_L$, shows a 3.7$\sigma$ combined discrepancy with the SM predictions in the region $4<q^2<8\gevgevcccc$. For $\Bs\to \phi \mu^+\mu^-$ and $\Lb\to \Lz \mu^+\mu^-$, the mismatch is seen for the differential branching fraction $d\mathcal{B}/dq^2$ in the region $q^2<8\gevgevcccc$. All these trends will be followed up with more statistics. For the time being, no sign of deviation is observed in any of the observables for $\Bu\to \pip \mu^+\mu^-$.
\section{Semileptonics}
Semileptonic decays $b\to (c,u)\ell^-\overline\nu$ have the advantage of providing an exact factorization between the weak part of the decay $\ell^-\overline\nu$ and the hadronic $b\to (c,u)$ transition involving form factors of the type $b$-hadron ($H_b$) $\to$ open-charm ($H_c$) or light-quarks hadron ($H_q$), $F(H_b\to H_{c,q})(q^2)$, where $q^2=(\tilde p_b-\tilde p_{c,q})^2$. The two parts are connected through a $W$ boson in the SM with an intervention of the CKM elements $V_{cb}$ and $V_{ub}$. 

\begin{figure}[htb]
\begin{center}
\includegraphics[width=0.3\textwidth]{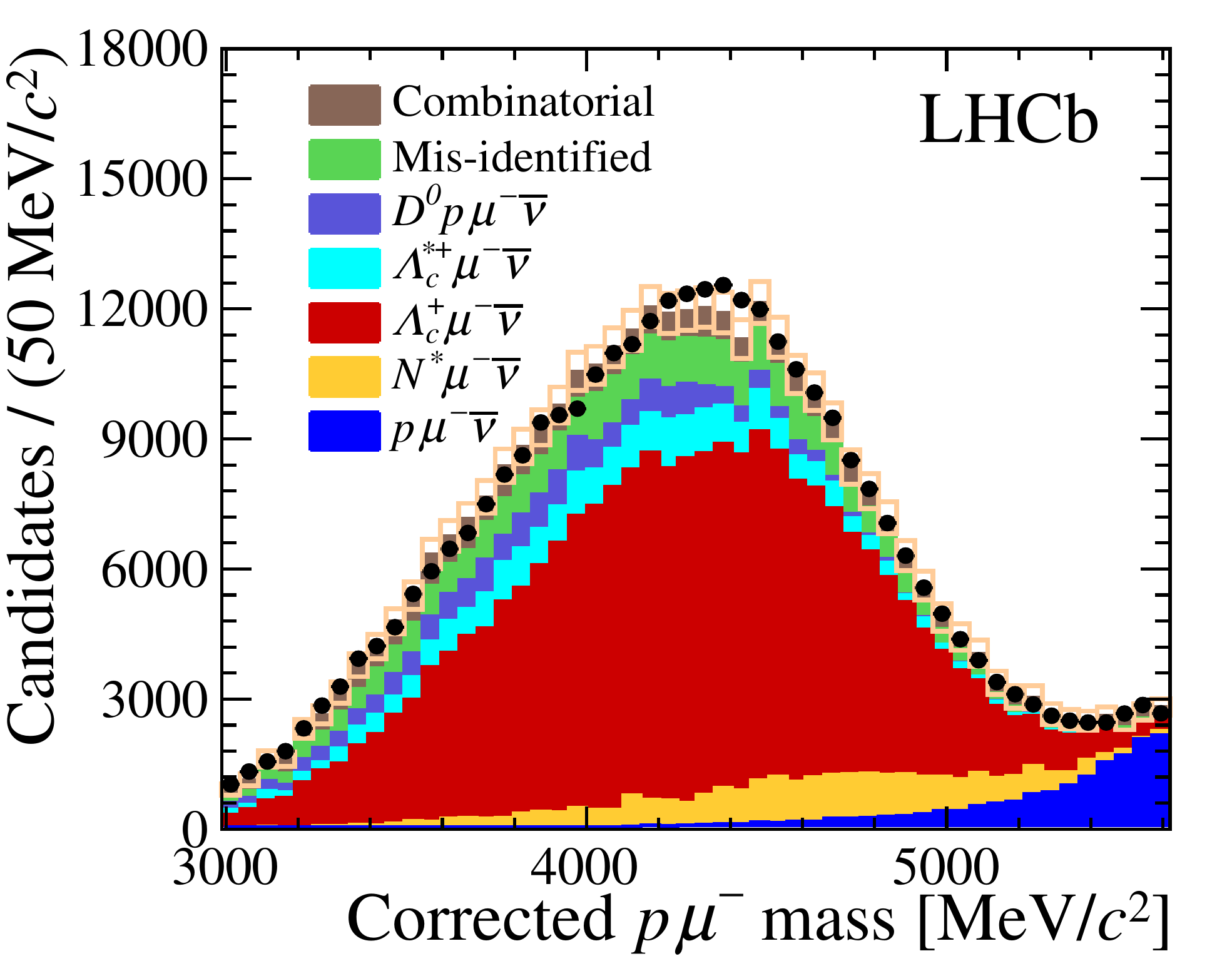}
\includegraphics[width=0.3\textwidth]{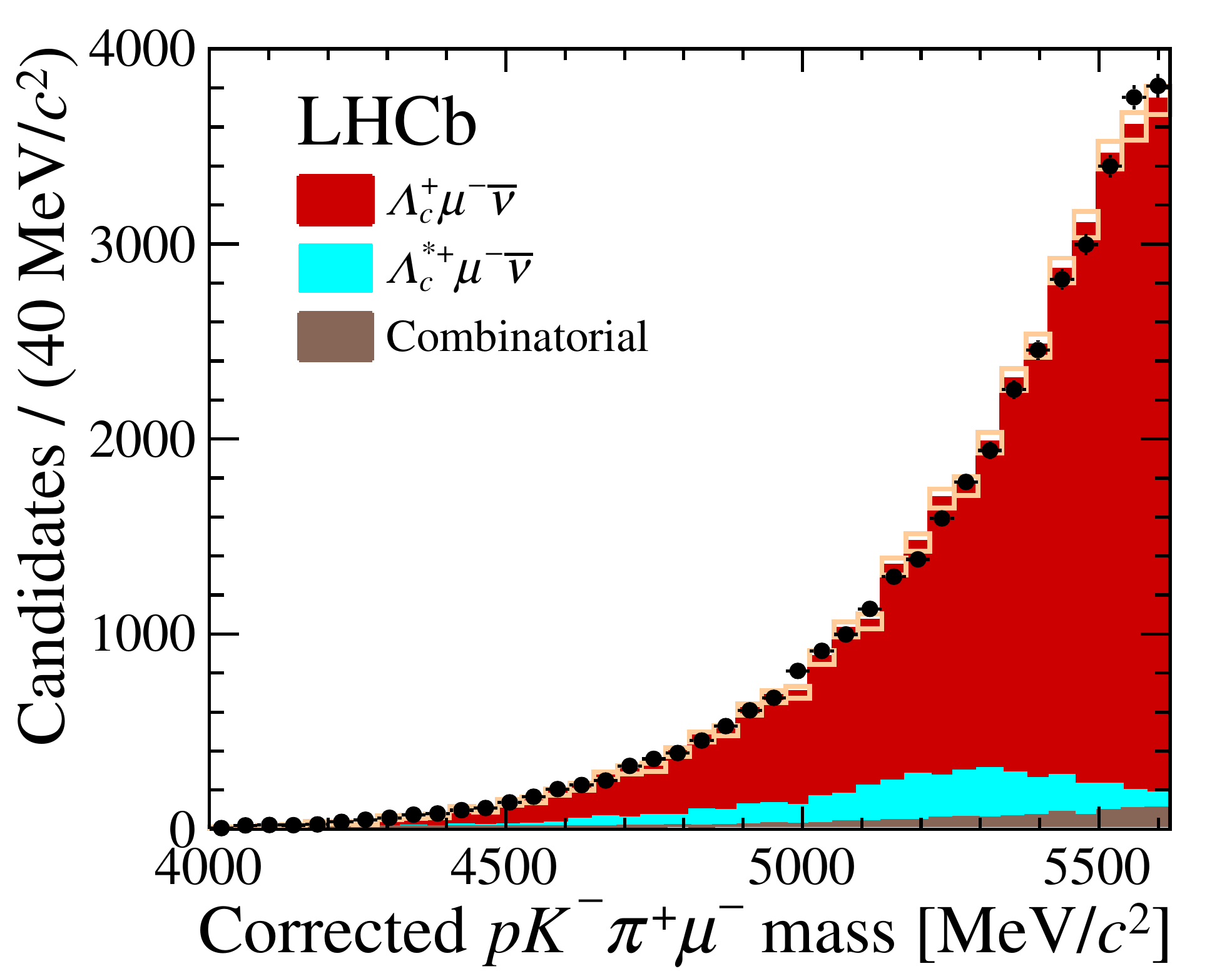}
\end{center}
\caption{Corrected mass distributions of the (top) $\Lb\to\proton\mu\overline \nu$ and (bottom) $\Lb\to\Lc(\to pK\pi)\mu\overline \nu$ candidates, showing contributions of the signals and backgrounds.}
\label{Fig:Mcorr_Lb}
\end{figure}

$V_{cb}\sim 4\times 10^{-2}$ is known to a typical relative accuracy of 2\% while the precision on $V_{ub}\sim[3-4]\times 10^{-3}$ is not better than 12\% \cite{PDG2014_SL}. Measuring $V_{ub}$ is thus interesting in itself. It becomes even more attractive with the persistence of a 3$\sigma$ discrepancy between measurements from inclusive $B\to X_u\ell\overline \nu$ and exclusive $B\to\pi\ell\overline \nu$ decays, with $(V_{ub})_{incl}>(V_{ub})_{excl}$. The pattern $(V_{cb})_{incl}>(V_{cb})_{excl}$ is also observed, though not as significant. One explanation that has been put forward \cite{VubTh} is the possible existence of a right-handed current brought by extra operators generated by a TeV-scale new physics. \lhcb has investigated the baryonic equivalent of $B\to\pi\mu\overline \nu$, $\Lb\to\proton\mu\overline \nu$ \cite{LHCb-PAPER-2015-013}, using $\Lb\to\Lc(\to pK\pi)\mu\overline \nu$ as a control channel. As there is no access to the missing neutrino energy, $q^2$ is inferred from the \Lb flight vector, the visible momentum, with a \Lb mass constraint. The discriminating variable is the corrected mass, $M_{corr}=\sqrt{M_{vis}^2+p_\perp^2}+p_\perp$ where $p_\perp$ is inferred from the flight vector and the visible momentum. $M_{vis}=M_{p\mu}$ for $\Lb\to\proton\mu\overline \nu$ and $M_{vis}=M_{pK\pi\mu}$ for $\Lb\to\Lc(\to pK\pi)\mu\overline \nu$. The distribution of this variable is shown in Fig.\ref{Fig:Mcorr_Lb} for both channels.
From the efficiency-corrected $q^2$-integrated yields and the use of $\Lb\to\proton$ and $\Lb\to\Lc$ form factors calculated with lattice QCD \cite{pmunuTh}, one can infer $|V_{ub}|/|V_{cb}|=0.083\pm0.004(\mathrm{exp})\pm0.004(\mathrm{lattice})$. This gives finally: $|V_{ub}|=(3.27\pm0.15(\mathrm{exp})\pm0.16(\mathrm{lattice})\pm0.06(|V_{cb}|))\times 10^{-3}$.

\begin{figure}[htb]
\begin{center}
\includegraphics[width=0.35\textwidth]{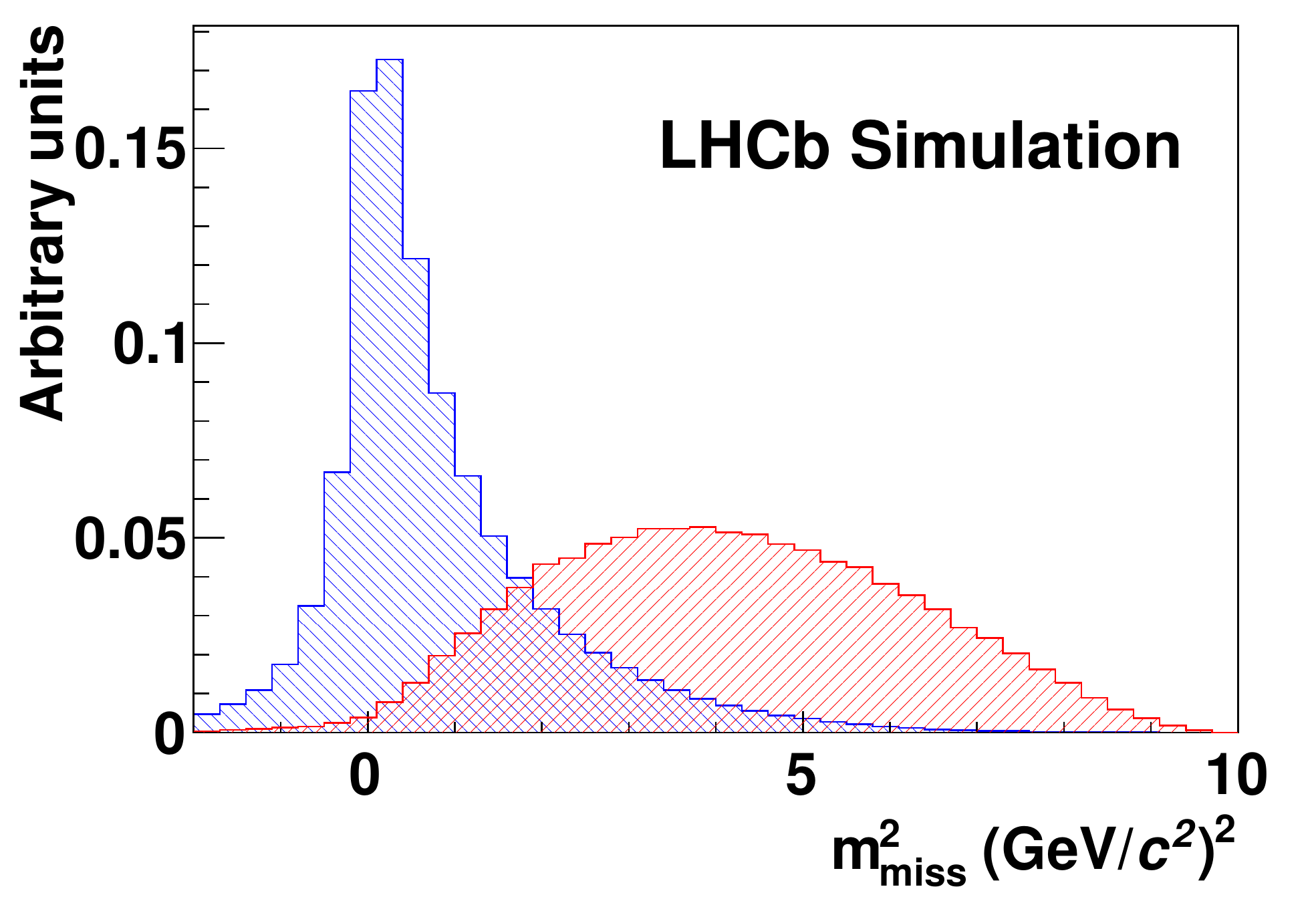}
\end{center}
\caption{Squared missing mass distributions for (blue) $B\to D^*\mu^-\overline\nu_\mu$ and (red) $B\to D^*\tau^-\overline\nu_\tau$ reconstructed events from simulation.}
\label{Fig:mmiss_dstlnu}
\end{figure}

\begin{figure}[htb]
\begin{center}
\includegraphics[width=0.35\textwidth]{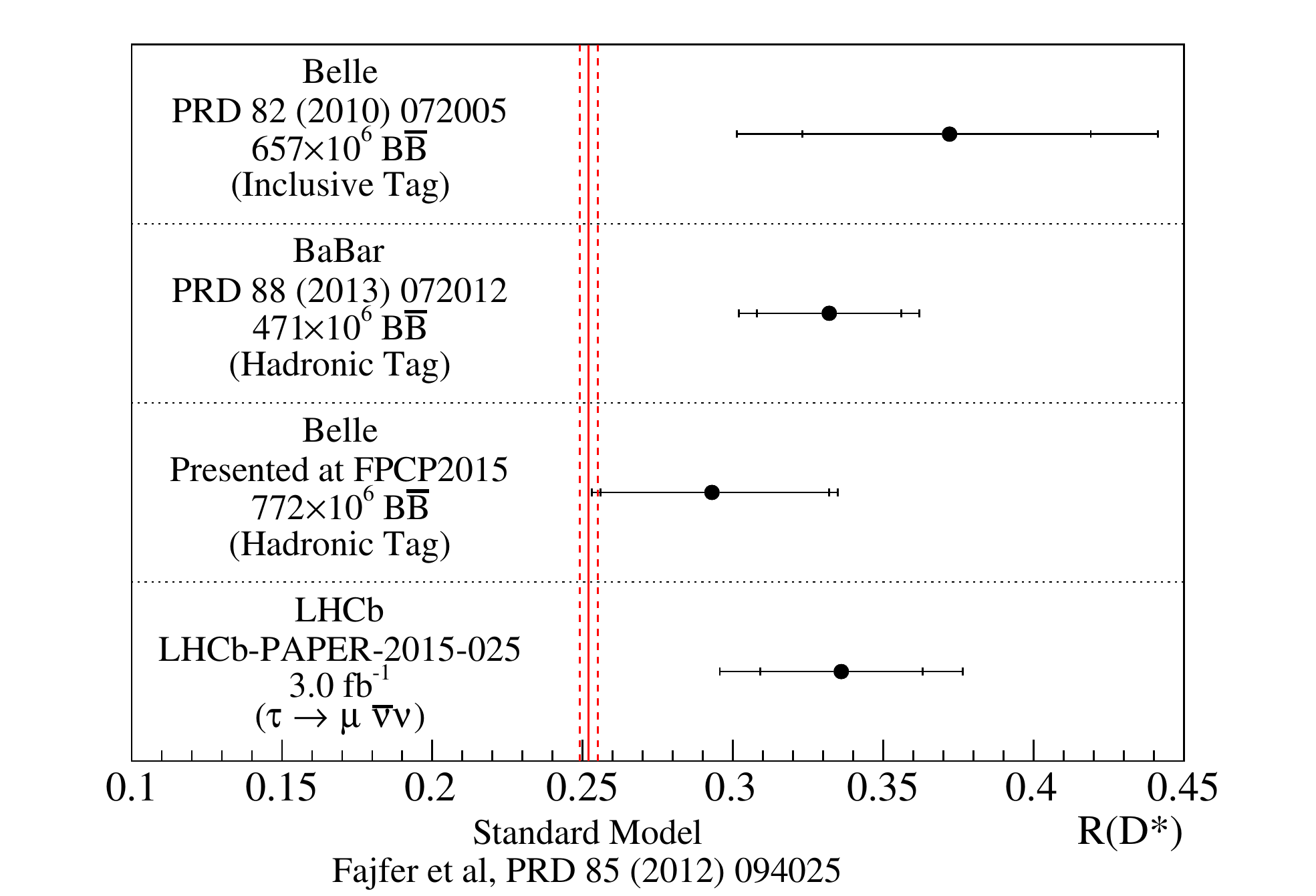}
\end{center}
\caption{Comparison of results on $R(D^*)$ between past and current measurements, and SM-based expectations.}
\label{Fig:RDstar_comp}
\end{figure}

Any beyond-SM heavy charged particle replacing the $W$ in $W\to\ell^-\overline\nu$ would preferably couple to a $\tau$ lepton. Therefore, by studying e.g. the ratio of $b\to c\tau^-\overline\nu_\tau$ to $b\to c\mu^-\overline\nu_\mu$, a deviation induced by new physics could be probed. This is done in \lhcb by measuring $R(D^*)=\mathcal{B}(B\to D^*\tau^-\overline\nu_\tau)/\mathcal{B}(B\to D^*\mu^-\overline\nu_\mu)$ \cite{LHCb-PAPER-2015-025}. The $D^*$ is reconstructed in the chain $D^{*+}\to D^0(\to K\pi)\pip$ while the $\tau$ is considered in the decay $\tau^-\to\mu^-\overline\nu_\mu\nu_\tau$ to allow for the same final visible particles for both $D^*\tau^-\overline\nu_\tau$ and $D^*\mu^-\overline\nu_\mu$. Apart from the the variable $q$, the muon energy and the invariant mass of the invisible part of the decay are used as discriminating variables. The latter one is important to separate $D^*\tau^-\overline\nu_\tau$ which has a big invisible mass from $D^*\mu^-\overline\nu_\mu$ for which this quantity is much smaller as can be seen in Fig.\ref{Fig:mmiss_dstlnu}.
The fits to the two channels, accounting for the corrections to the detection and reconstruction efficiencies, permit to extract their ratio as $R(D^*)=0.336\pm0.027(\mathrm{stat})\pm0.030(\mathrm{syst})$, which confirms (+2.1$\sigma$) the trend observed in the past above the SM-based predictions \cite{RDst_Th}, as shown in Fig.\ref{Fig:RDstar_comp}.
\section{CKM $\gamma$ angle}
The easiest way to extract the angle $\gamma$ is to use the interference of tree favored $b\to c (\overline u s)$ ($A_B$) and suppressed $b\to u (\overline c s)$ ($A_Br_B e^{i(\delta_B-\gamma)}$, $r_B$ is the suppression factor and $\delta_B$ is the relative strong phase) decays leading to final states of the type $B\to D X_s$ where $D$ is a charm meson and $X_s$ a strange system ($K$, $K^*$, $K\pi$, $K\pi\pi\pi$, ...). The charm meson decays to a final state accessible by both $D$ and $\overline D$ through allowed ($A_D$) and suppressed ($A_Dr_De^{i\delta_D}$). Since more than two decades, several methods have initially been proposed for $B\to D K$ \cite{GLW,ADS,GGSZ,GLS} to extract the angle for observables including charge asymmetries and combination of branching fractions which definition may vary depending on the type of $D$ decay (\CP eigenstate, flavour eigenstate, multibody). Beside the generalization of the paradigm to more complex $X_s$ system, alternative methods have been developped with other $B$ mesons, such as $B^0_s \to D^\mp_s K^\pm$ \cite{LHCb-PAPER-2014-038}. 

Over the last years, the \lhcb collaboration has been combining several $B\to DK$ modes to extract a value of $\gamma=(72.9^{+9.2}_{-9.9})^\circ$ \cite{LHCb-CONF-2014-004}.
For multibody $D$ decays, the variation of the strong phase $\delta_D$ across the decay phase space engenders a dilution factor in the extraction of the $\gamma$ angle. This effect can be unfolded by dedicated amplitude analyses of this type of decays. As an illustration, \lhcb used a recent determination of the \CP content of $D^0\to\pip\pim\piz$ and $D^0\to\Kp\Km\piz$ based on CLEO-c data \cite{cleoc_dhhpi0} to perform an extraction of $r_B$, $\gamma$ and $\delta_B$ in $B\to D(\to h^+h^-\piz)K$ decays \cite{LHCb-PAPER-2015-014}. The accuracy of the results is not yet at the level of giving a substantial contribution to the combined fit. 

The other recent study involves a $B\to D X_s$ decay, $B\to D(\to hh) K\pi\pi$ \cite{LHCb-PAPER-2015-020}. The two observables used to extract the relevant information are charge averages based on the decay rates, $\mathcal{B}(\Bpm\to D(\to hh)X_s^\pm)\propto 1+r_B^2+2\kappa r_B\cos(\delta_B\pm\gamma)$, and the charge asymmetry $\mathcal{A}(\Bpm\to D(\to hh)X_s^\pm)\propto 2\kappa r_B \sin\delta_B\sin\gamma$, where $\kappa$ is a dilution factor due to the multibody nature of the decay $B\to D X_s$. From simultaneous fits to the signals $B\to D(\to K\pi) X_s$, $B\to D(\to \pi\pi) X_s$, $B\to D(\to KK) X_s$, the value $\gamma=(74^{+20}_{-18})^\circ$ is obtained, in agreement with the combined average.
\section{New quark bound states}
\begin{figure}[htb]
\begin{center}
\includegraphics[width=0.35\textwidth]{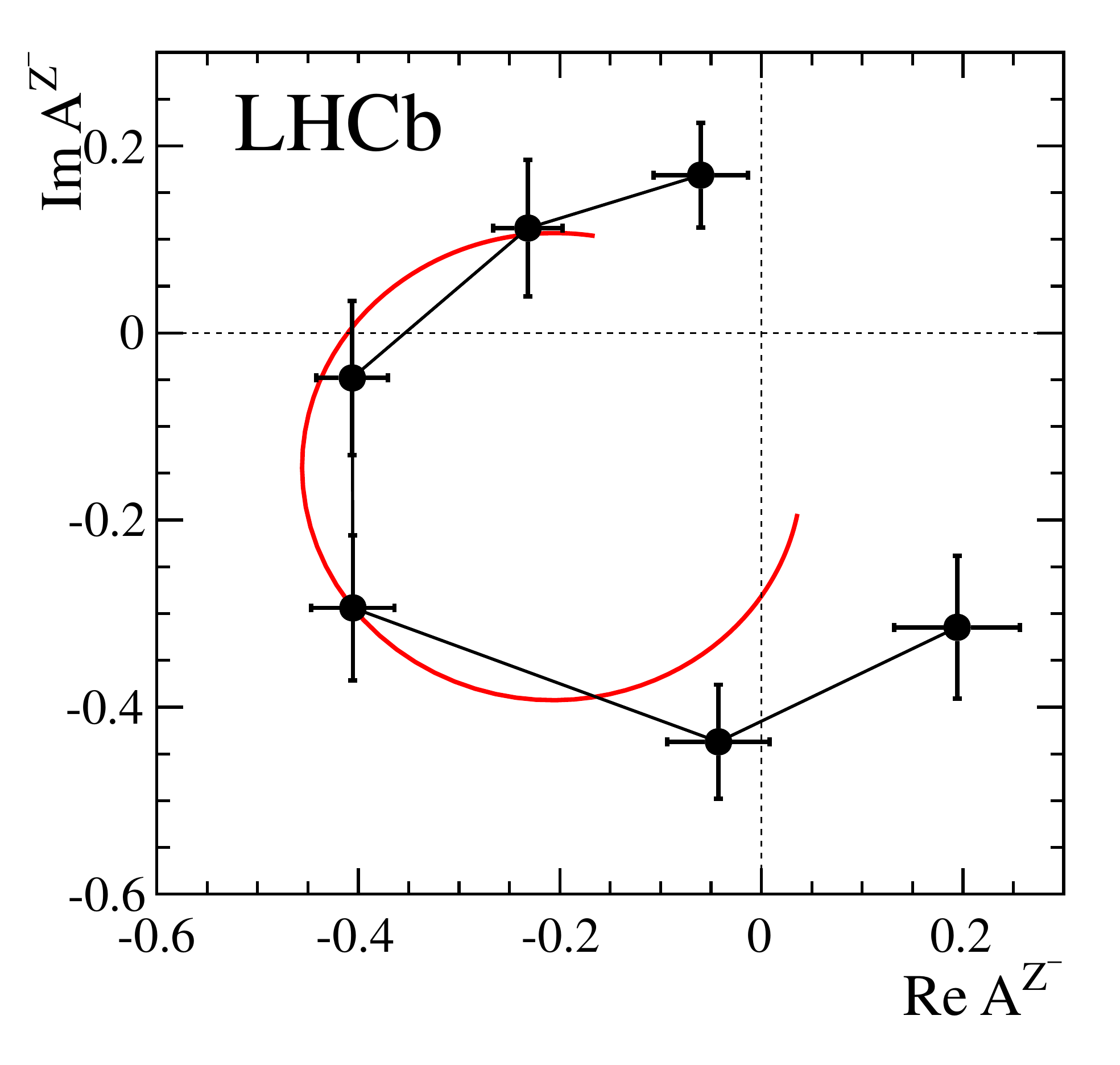}
\end{center}
\caption{Fitted value of the $Z(4430)^-$ amplitude in six $m^2(\psi^\prime\pim)$ bins (increasing counterclockwise). The red curve shows the prediction from a Breit-Wigner formula with mass and width of 4475 and 172 \mevcc, respectively.}
\label{Fig:ZArgand}
\end{figure}

The idea of hadronic bound states having more than three valence quarks has been envisaged since already half a century \cite{gellmann}, with subsequent quantitative work including tetraquark and pentaquark states \cite{jaffe,strottman,lipkin}. On the experimental side, various past claims of pentaquark observation have been shown to rely on unconvincing signals \cite{hicks}. Recently however, the resonant nature of the tetraquark candidate $Z(4430)^-$ seen in the $\psi^\prime\pim$ invariant mass distribution, first advertised by the Belle collaboration in $B\to K\psi^\prime\pi$ decays \cite{belletetra1,belletetra2}, has been finally established by \lhcb \cite{LHCb-PAPER-2014-014}. The Argand diagram of the amplitude is shown in Fig.\ref{Fig:ZArgand}. Current independent studies aim at consolidating the result \cite{LHCb-PAPER-2015-038}.

\begin{figure}[htb]
\begin{center}
\includegraphics[width=0.3\textwidth]{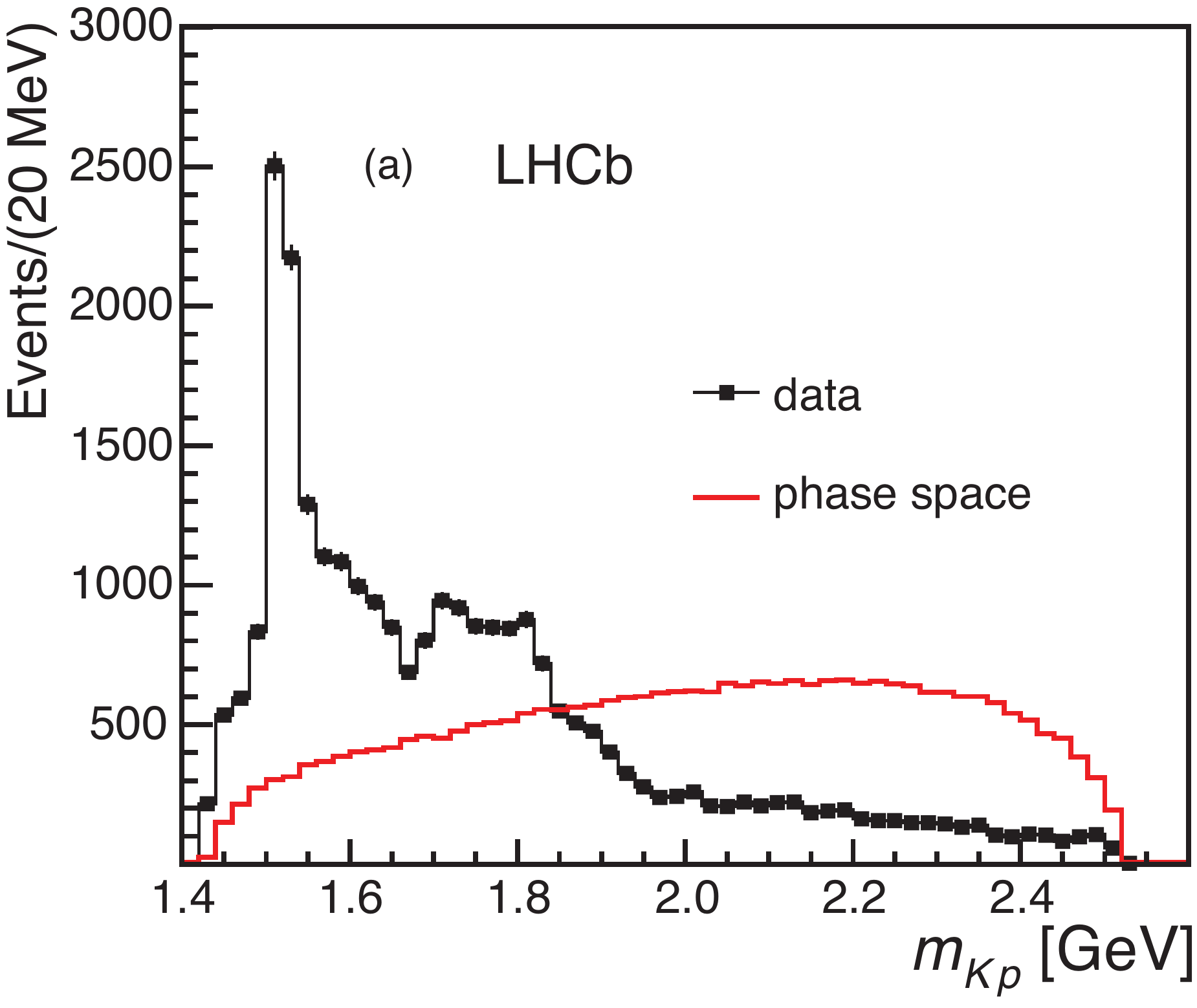}
\includegraphics[width=0.3\textwidth]{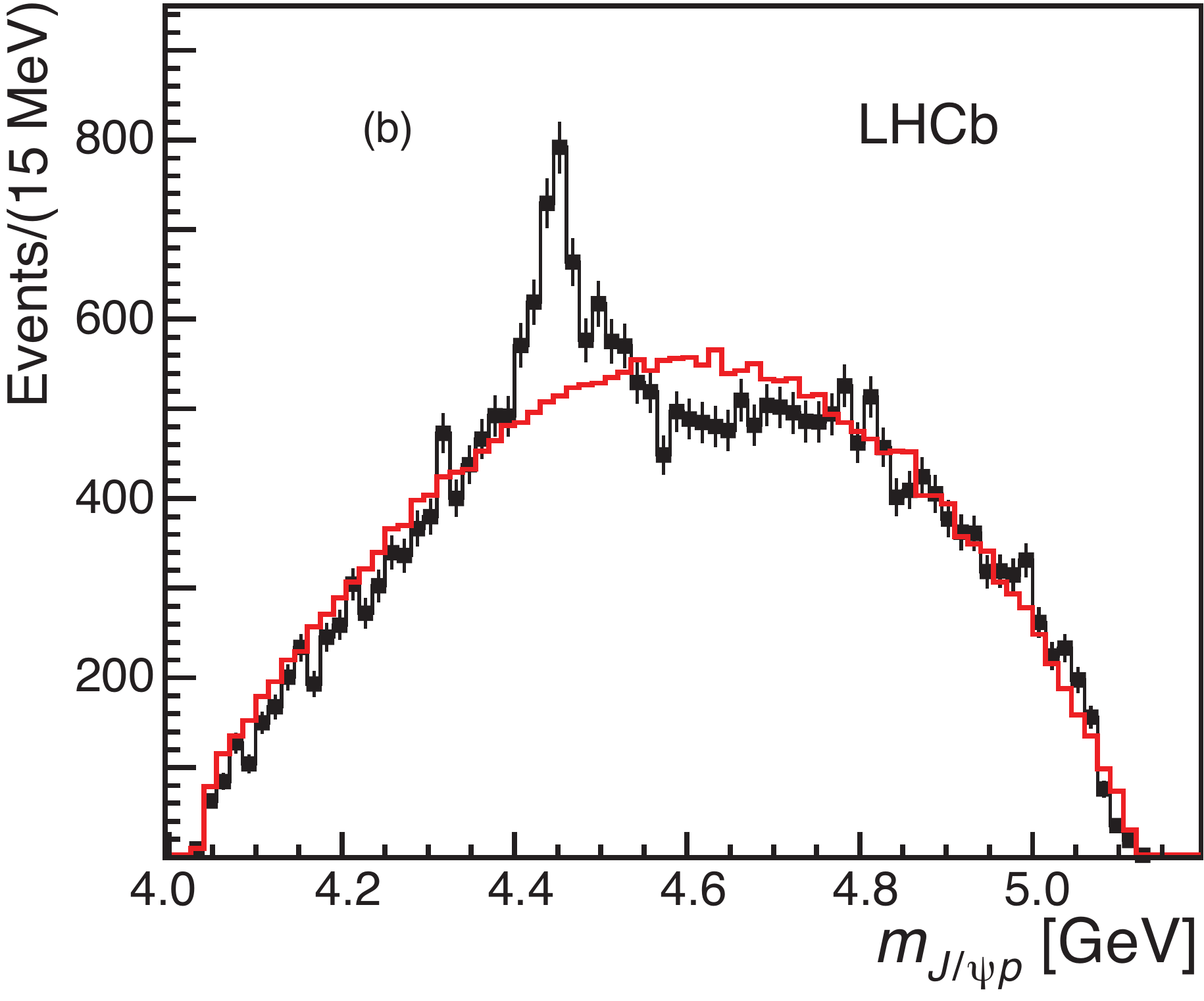}
\end{center}
\caption{Invariant masses of (top) $Kp$ and (bottom) $\jpsi\proton$ combinations for the signal decays $\Lb\to\jpsi\proton K$. The peaking structure above 4.4\gevcc in $m(\jpsi\proton)$ is clearly visible. The complex $Kp$ pattern below 2.0\gevcc is produced by combinations (and interferences) of $\Lambda$ resonances.}
\label{Fig:LbjpsiKp_masses}
\end{figure}

\begin{figure}[htb]
\begin{center}
\includegraphics[width=0.45\textwidth]{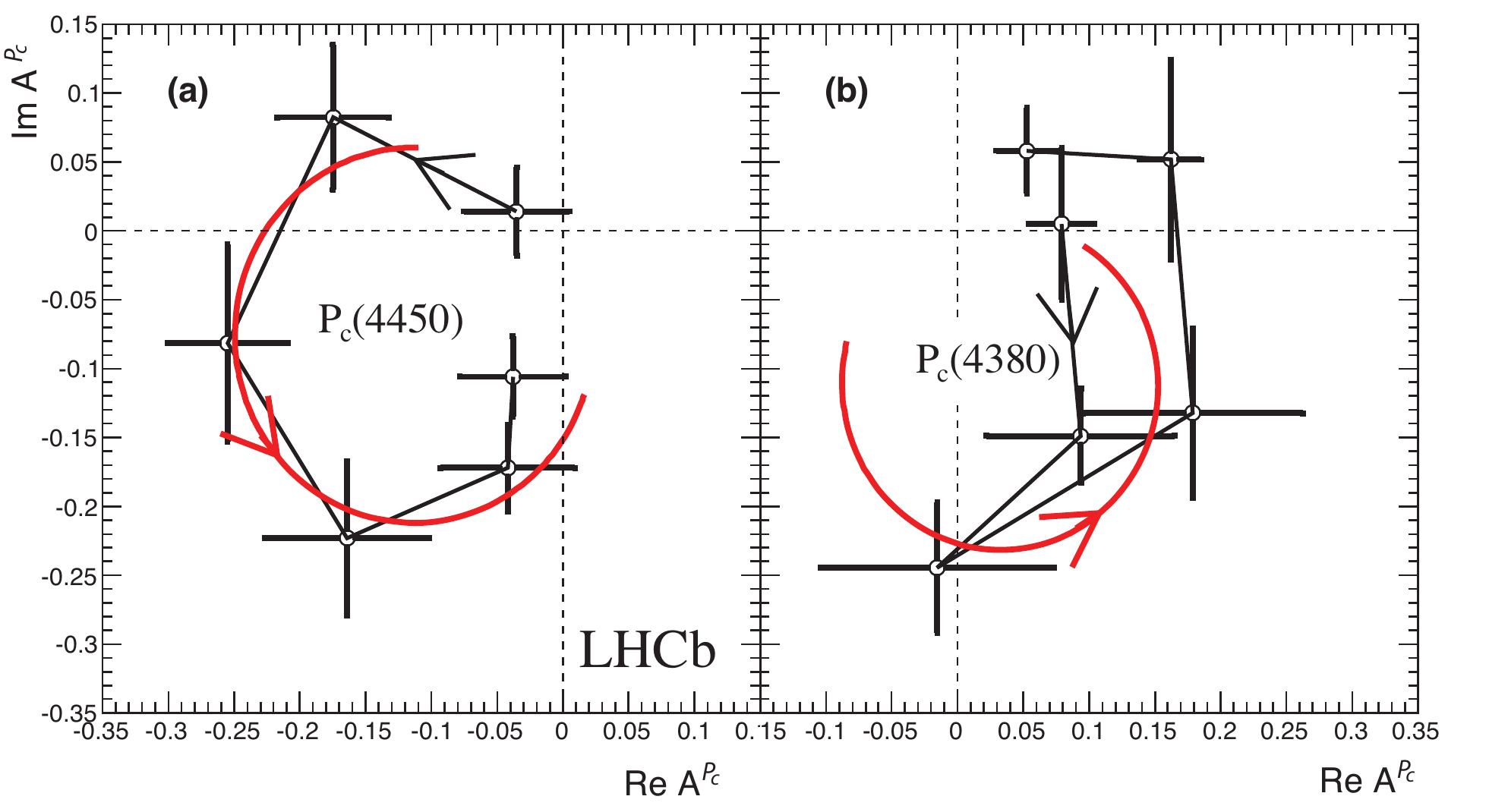}
\end{center}
\caption{Fitted value of the (a) $P_c^+(4450)$ and (b) $P_c^+(4380)$ amplitudes in $m^2(\jpsi\proton)$ bins (increasing counterclockwise). The red curve shows the predictions from relativistic Breit-Wigner formulaes with masses, widths and spins corresponding to values found in the amplitude analysis (see text).}
\label{Fig:Pc_Argand}
\end{figure}

While studying the baryonic counterpart of the aforementioned decays, which are $\Lb\to\jpsi\proton K$ decays, the \lhcb collaboration achieved a significant breakthrough in the path of proving the existence of a pentaquark resonance \cite{LHCb-PAPER-2015-029}. The $\jpsi\proton$ invariant mass spectrum exhibits a peculiar peak near 4.4\gevcc as shown in Fig.\ref{Fig:LbjpsiKp_masses}. A full amplitude analysis involving the squared invariant masses of the $\jpsi\proton$ and $K\proton$ systems, and five decay angles, has been performed to investigate the decay spectrum. The best fit leads to the data being compatible with the existence of two resonances $P_c^+(4380)$ and $P_c^+(4450)$ of masses $(4380\pm8\pm29,~4449.8\pm1.7\pm2.5$ \mevcc and widths $(205\pm18\pm86,~39\pm5\pm19$, respectively, and spin-parities equal to $(3/2^-,5/2^+)$. The signal of the narrower resonance is stronger as already shown in Fig.\ref{Fig:LbjpsiKp_masses} and less likely to be sensitive to fluctuation of other amplitudes. This is reflected in the Argand diagrams of Fig.\ref{Fig:Pc_Argand}.

Both $Z$ and $P_c$ are built on charmonium and are thus labeled as charmonium-tetraquark and charmonium-pentaquark states. The exact dynamics and structure of these states are still subject to intense modelings and investigations.

\section{Forward physics}
Over the past few years, there has been a significative number of \lhcb studies involving $W$ or $Z$ bosons, jets and searches for displaced vertices of long-lived heavy particles in the forward region \cite{LHCb-PAPER-2014-062,LHCb-PAPER-2015-002}. Since the first publications on inclusive electroweak bosons production \cite{LHCb-PAPER-2012-008,LHCb-PAPER-2014-033}, the collaboration has been particularly active in improving the quality of the identification of heavy-flavoured jets \cite{LHCb-PAPER-2015-016} with the idea of studying the boson-jet associated production \cite{LHCb-PAPER-2015-021}. 

\begin{figure}[htb]
\begin{center}
\includegraphics[width=0.3\textwidth]{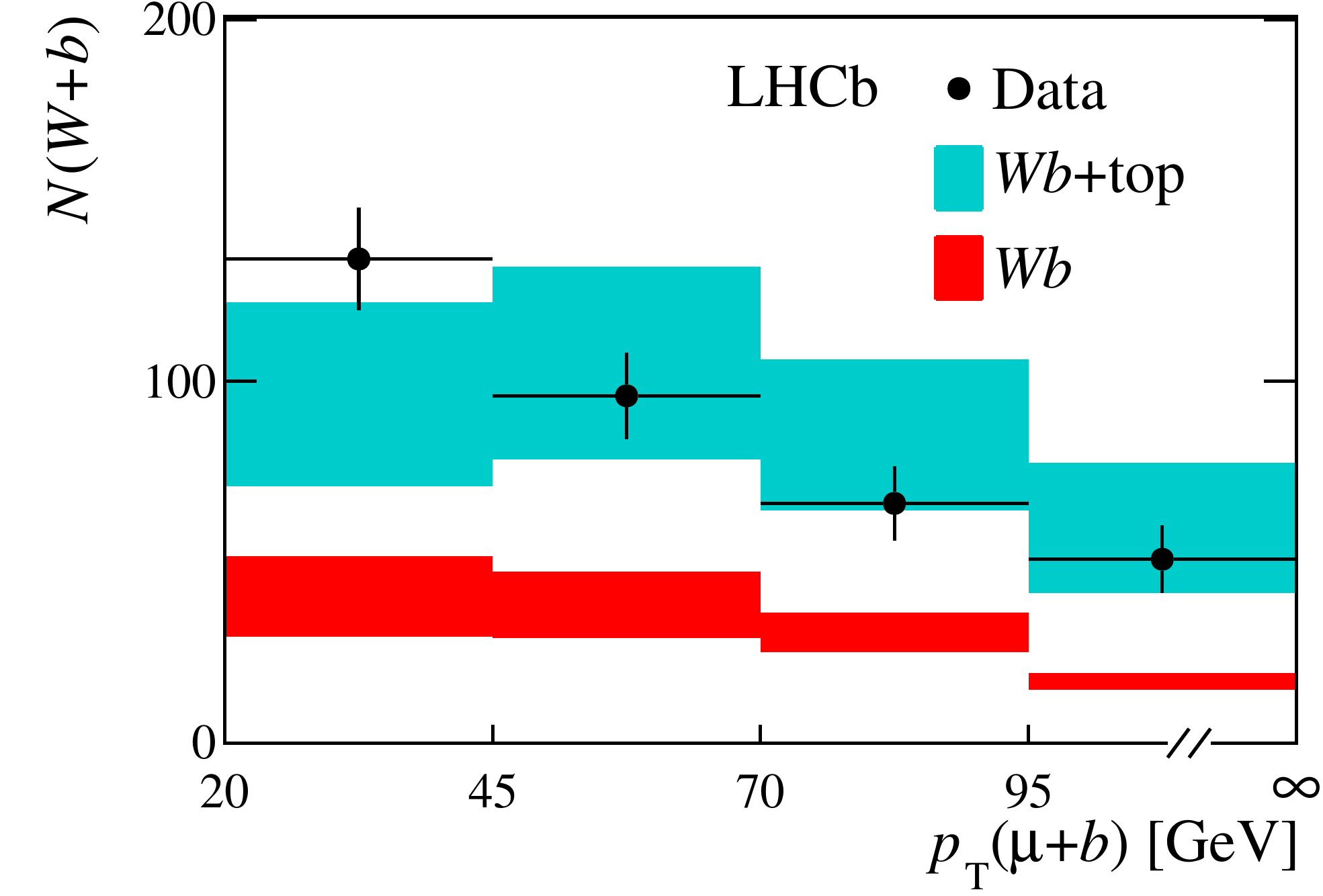}
\includegraphics[width=0.3\textwidth]{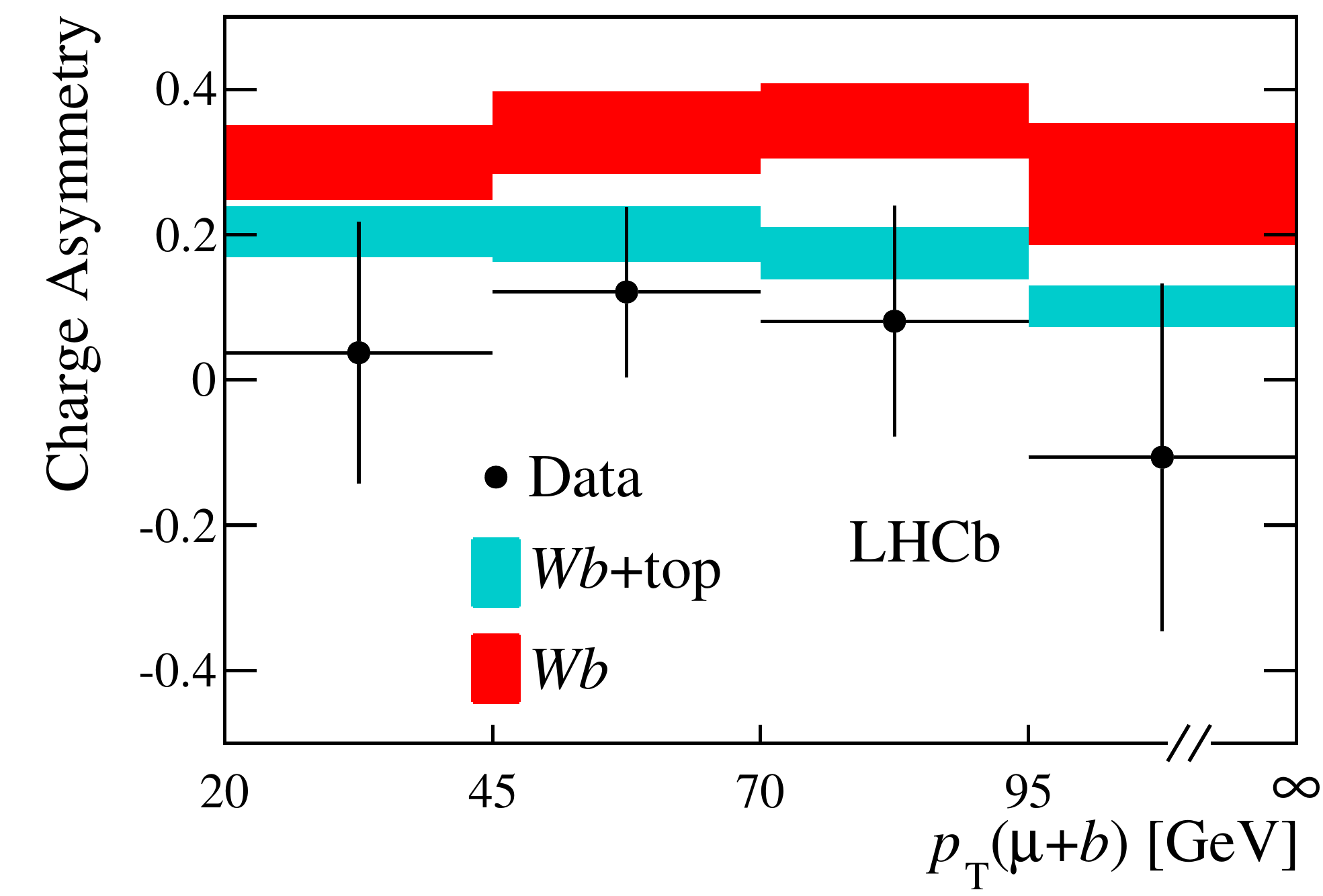}
\end{center}
\caption{Fit results for the (top) $W+b$ yield and (bottom) $W+b$ charge asymmetry in bins of $\pt(\mu+b)$, compared to NLO level SM predictions.}
\label{Fig:Wb_top}
\end{figure}

The search for top production in the $W+b$-jet associations comes as a natural follow-up of this effort. In reference \cite{LHCb-PAPER-2015-022}, the muon from the $W\to\mu\nu$ decay is combined with a jet, with further multivariate discrimination between $W+q$-jet ($q$ denotes light-quark jets), $W+c$-jet and $W+b$-jet categories. The inclusive $W+j$-jet production is used as a control channel, since it has very low background and the ratio $\sigma(Wb)/\sigma(Wj)$ is predicted to a very good accuracy in the Standard Model. Using the latter quantity also permits to benefit from the cancelling of several experimental uncertainties. The yields are extracted by a fit to the transverse momentum ratio $\pt(\mu)/\pt(j_\mu)$, where $j_\mu$ denotes the jet clustered around the muon, including the muon. This variable is particularly useful to reject dijet events. Figure \ref{Fig:Wb_top} shows the fit results for $W+b$ yield and charge asymmetry $(N(W^+b)-N(W^-b))/(N(W^+b)+N(W^-b))$ in bins of $\pt(\mu+b)$: the data favours the existence of the top production over pure $W+b$-jet production by a significance of 5.4$\sigma$, which establishes the observation. 

 For what concerns Higgs physics, a limit on Higgs decays to $\tau^+\tau^-$ has already been set by \lhcb with the 2011 data sample \cite{LHCb-PAPER-2013-009}. However, the experimental improvements that permitted the aforementioned results on $Wb$, $Wc$ and top identification have opened the door to exciting studies on boson + dijet production in the forward region with the hope of probing the boson + Higgs forward production. 

\section{Current data taking and prospects}
\begin{figure}[htb]
\begin{center}
\includegraphics[width=0.45\textwidth]{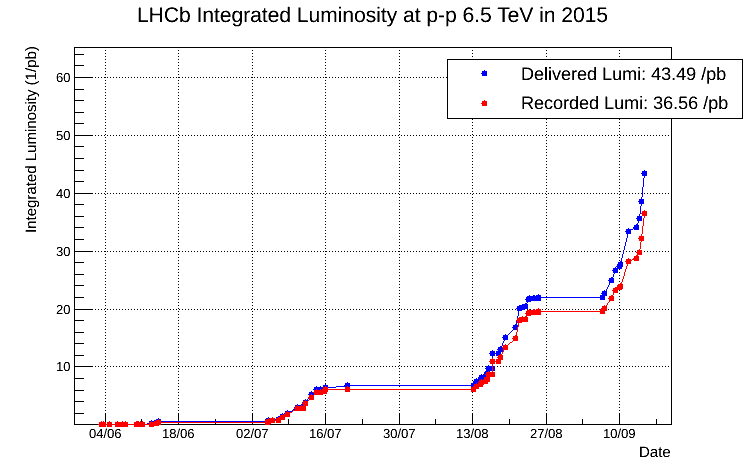}
\end{center}
\caption{Integrated luminosity as a function of time for the beginning of Run II data taking.}
\label{Fig:2015Lumi}
\end{figure}

\begin{figure}[htb]
\begin{center}
\includegraphics[width=0.3\textwidth]{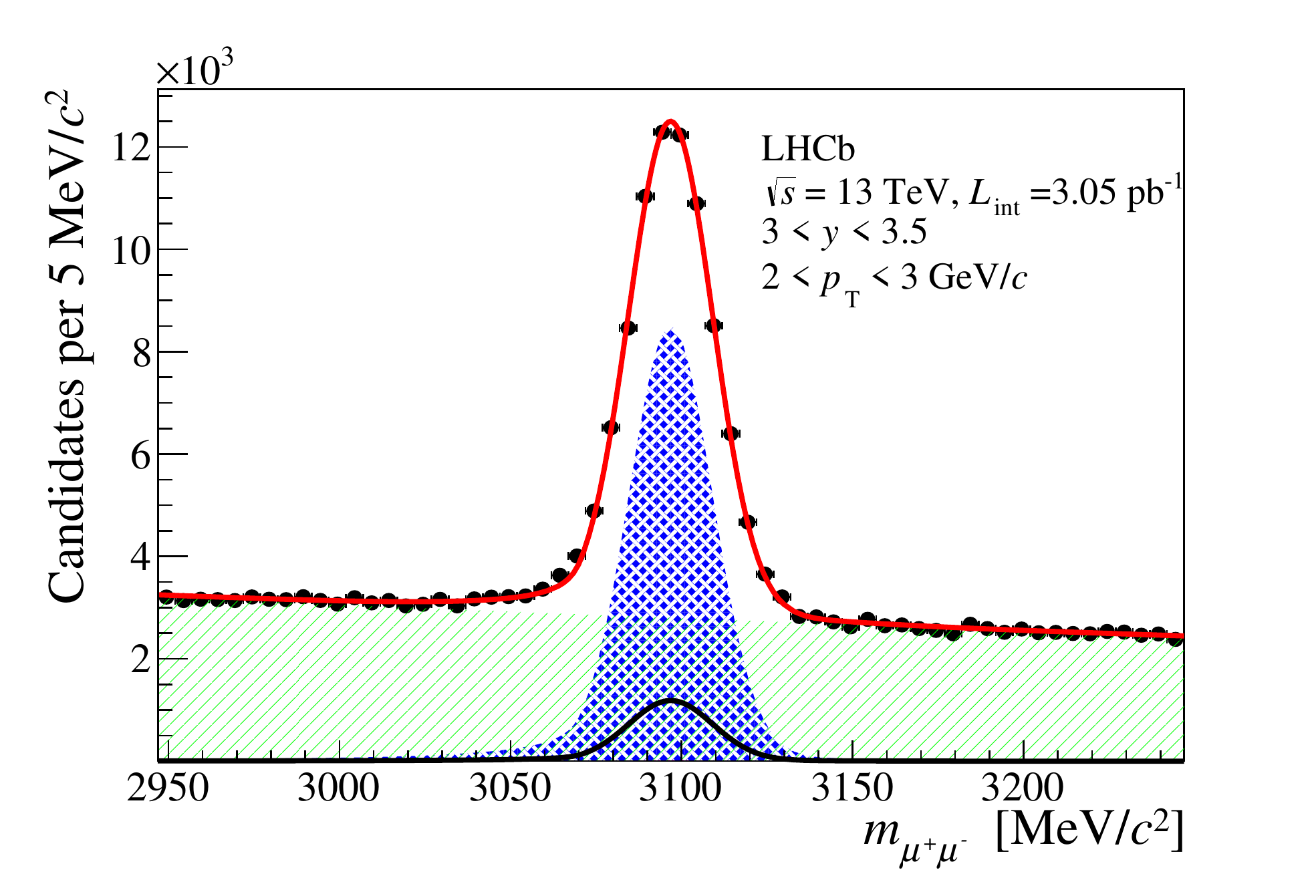}
\includegraphics[width=0.3\textwidth]{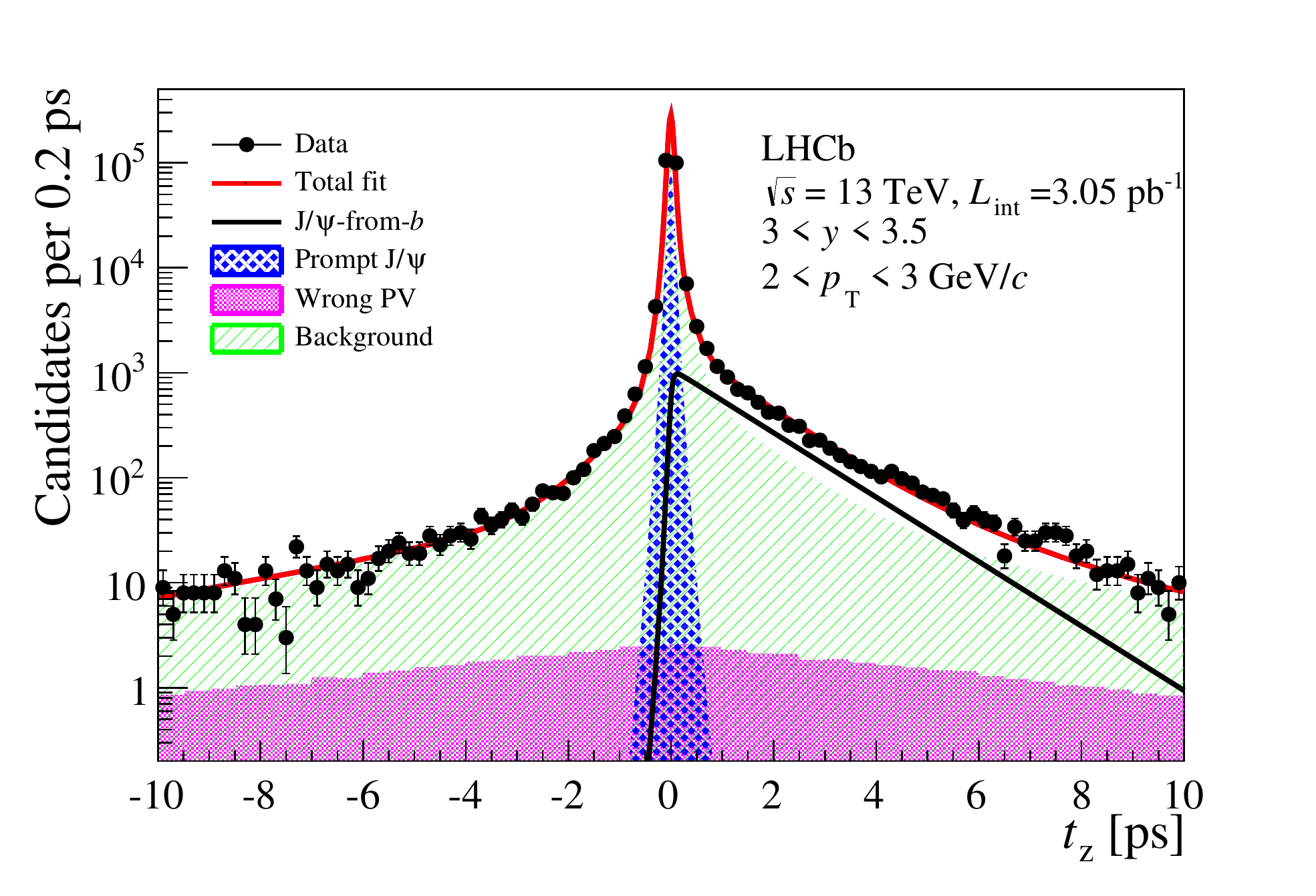}
\end{center}
\caption{$\mu\mu$ invariant mass (top) and pseudo-decay time (bottom) for the bin $2<\pt<3\gevc$, $3.0<y<3.5$. The prompt component is shown in blue cross-hatched area while the solid black line exhibits the \jpsi mesons coming from $b$-hadrons.}
\label{Fig:jpsi_early}
\end{figure}

Run II data taking has started early June 2015 at a center of mass energy of 13 TeV with a proton bunch spacing of 50 ns. Since early September, the 25 ns bunch spacing is used and the current status of the integrated luminosity is shown in Fig.\ref{Fig:2015Lumi}. The main difference in the data acquisition with Run I is the online calibration and alignment performed at the High Level Trigger step, and a larger output rate to storage (12.5 kHz). The 50 ns data sample has already given birth to several validation plots and studies, with the first early publication being on the \jpsi production \cite{LHCb-PAPER-2015-037} based on an integrated luminosity of 3\invpb. A fit to the $\mu\mu$ invariant mass and vertex lifetime distributions is performed to extract the prompt and \jpsi-from-$b$-hadron components, in bins of transverse momentum \pt and rapidity $y$. Figure \ref{Fig:jpsi_early} illustrates the fit results for one kinematical bin.

The cross-sections extracted in the \lhcb acceptance are $\sigma_{\mathrm{prompt}\jpsi}(\pt<14\gevc, 2.0<y<4.5)=15.30\pm0.03\pm0.86\mub$ and $\sigma_{\jpsi-\mathrm{from}-b}(\pt<14\gevc, 2.0<y<4.5)=2.34\pm0.01\pm0.13\mub$, where the firt and second uncertainties are statistical and systematic, respectively. Using the $b\to\jpsi X$ branching fraction \cite{PDG2014} and an acceptance extrapolating factor from \pythia simulation, the total cross-section is obtained $\sigma(pp\to b\overline b X)=515\pm2\pm53\mub$.

For Run II, the benchmark scenario predicts an integrated luminosity of $\approx$ 5-6 \invfb by 2018, which will more than double the Run I sample, given the $b\overline b$ production cross-section scaling with energy. This is will certainly allow to converge on a final word for all the deviations seen for several observables in the rare $B$ decays. In the case where beyond-SM physics does not show up directly, the indirect presence of new physics would permit to estimate its energy scale through the size of its contribution to low-energy effective hamiltonians. Besides, the higher center-of-mass energy of Run II (and further data taking) enhances the capabilities of multiple-jets detection in \lhcb and thus improves in particular the sensitivity to associated Higgs and electroweak bosons production.

\section{Conclusion}
The harvest of physics results for Run I data has permitted to drastically improve the knowledge of the weak \CP violation in the quark sector with in particular measurements of the $\beta_s$ and $\gamma$ angles. From the improvements on rare decays, it has also shed light on the size of possible contributions of new heavy particles in loops, with the remarkable observation of $B\to\mup\mu^-$\cite{LHCb-PAPER-2014-049}. The \lhcb collaboration has also shown the potential for interesting prospects for electroweak physics in the forward region with the first observation of the top quark production.

The upgrade of years 2019-2020 \footnote{See dedicated presentation, these proceedings} is aimed at preparing the detector for the high luminosity run, where about 50 \invfb of data are expected to be recorded, which will project the collaboration into a new era of discoveries and precision tests for rare decays, heavy hadrons, \CP violation and forward physics.
\clearpage
\addcontentsline{toc}{section}{References}
\setboolean{inbibliography}{true}
\bibliographystyle{LHCb}
\bibliography{mybib,LHCb-PAPER,LHCb-CONF,LHCb-DP}

\end{document}